\documentclass[fleqn,usenatbib]{mnras}

\usepackage{newtxtext,newtxmath}

\usepackage[T1]{fontenc}
\usepackage{ae,aecompl}
\usepackage[utf8]{inputenc}

\usepackage{graphicx}	
\usepackage{amsmath}	

\title{}
\title[Counterjet lags]{The simultaneity of emission from approaching and receding jets}

\author[Maccarone, Pattie, \& Tetarenko]{
Thomas J. Maccarone$^1$, Eli C. Pattie$^1$ and Alexandra J. Tetarenko$^{1,2}$\\
$^1$Department of Physics \& Astronomy, Texas Tech University, Lubbock TX, 79410-1051, USA\\
$^2$ NASA Einstein Fellow
}

\date{Accepted XXX. Received YYY; in original form ZZZ}

\pubyear{2022}

\begin{document}
\label{firstpage}
\pagerange{\pageref{firstpage}--\pageref{lastpage}}
 \maketitle

\begin{abstract}
We show that the standard Blandford-K\"onigl model for compact conical relativistic jets has a peculiar feature: at a given observed frequency of radiation, the emission from the approaching jet arrives at the location of a distant observer at the same time as the emission from the counterjet for all finite inclination angles.  We show that this result can be used to determine whether jets are genuinely symmetric, if the cross-coherence between radio and X-ray time series can be measured at high Fourier frequency for a sample of neutron star X-ray binaries with a range of inclination angles.  We also discuss echo mapping techniques that can be used to look for deviations from the standard model in high cadence time series data on X-ray binary jets, and conclude that these can plausibly be applied to some systems.
\end{abstract}

\begin{keywords}
Black hole physics -- X-ray: binaries 
\end{keywords}



\section{Introduction}
Accreting black holes  and neutron stars often show flat (i.e. approximately constant flux per unit frequency) radio spectra.  These are traditionally explained as coming from compact conical jets.  If the random energy lost by the jet's electrons is replaced continuously, the wavelength at which the jet becomes transparent to synchrotron self-absorption is linearly proportional to the height up the jet, and the flat spectrum results \citep{BK79,HjellmingJohnston}.  While this approach has successfully explained many
of the properties of the radio emission from active galactic nuclei (AGN) and X-ray binaries, direct tests of the model are relatively few.  For AGN, the shifts in the positions of the cores of a modest-sized sample of objects observed with the Very Long Baseline Array are consistent with the model \citep{Sokolovsky2011}; these data confirm the general geometric picture, but do not give information about how the particle acceleration takes place.  For X-ray binaries, only the very nearest and most powerful jet sources are resolvable along the jet axis, and then barely so \citep{Fomalont,Stirling}, and are not resolvable perpendicular to the jet axis.  

Over the past decade, and especially in the past few years, attempts have been made to use variability information both to supplement the mapping of the parameter space from the core shifts\citep{ZdziarskiTetarenkoSikora}, and to try to understand the particle acceleration process \citep{Casella2010,Gandhi2017, Tetarenko2019,Tetarenko2021,Vincentelli2021}.  The results, particularly in \citet{Tetarenko2021}, are in qualitative, but not quantitative agreement with the standard model of \citet{BK79}, as the time lags are not linearly proportional to the wavelength.  In Cyg X-1, the deviations may occur due to the opacity of the stellar wind \citep{JamesCyg}.  In MAXI~J1820+070, the deviations are present, but modest, and could be due to acceleration in the jet, and/or changes in the jet opening angle.

When lags are measured using cross-correlation functions, the correlation coefficients at the peaks are typically much less than unity.  Thus, there is a ``typical'' lag, but the lags vary.  A theoretical model from \citet{Malzac2018} can explain this due to internal shocking, in agreement with the idea that the acceleration of particles in jets comes from this process \citep{Spada2001,Jamil}. Because of the incoherence of the lags, a large number of characteristic timescales of the variability must be sampled to measure the lags accurately.  This, combined with the much longer timescales of variability for AGN relative to X-ray binaries, means that X-ray binaries are the preferred class of systems for studying the lags, and hence for studying the acceleration process.  

A few other properties of the X-ray binaries make them more desirable targets for these studies.  The masses of the compact objects in X-ray binaries are typically (although not always) better measured than the masses of supermassive black holes.  The inclination angles of the X-ray binaries are also usually better measured, with the caveat that the the inclination angles that are measured are predominantly from the binary orbits, and there can be cases where the orbital and jet axes are misaligned in the long term (e.g. \citealt{Fragile2001,Maccarone20002,Miller-Jones2019,Poutanen2022}) and where the jet axes vary with time (e.g. \citealt{Milgrom1979,Tetarenko2017}).  Furthermore, neutron star X-ray binaries have jets which show a broad set of similarities to the jets from black hole X-ray binaries, but where their speeds have been measured, they tend to be significantly slower than the speed of light, and typically close to the expected escape velocities from neutron stars \citep{Fomalont}.  For these slower jets, the expected ratio of fluxes from the approaching jet to those from the counterjet due to relativistic Doppler boosting are typically less than 10 { (see Figure \ref{ratiofig})}, meaning that the signatures of the counterjets are likely to be much more prominent than they are for black hole X-ray binaries.

\section{Counterjet lags}
We start from the \citep{BK79} model for jets.  In this model, the height $h$ along the jet from which emission at wavelength $\lambda$ is emitted follows the relation $h\propto\lambda$.  We then define $h'$ to be the height along the counterjet where emission is produced at the same frequency.  The relativistic doppler factors for the approaching and counterjet will be $\delta_{\rm app}$ and $\delta_{\rm cj}$, respectively, and will be given by:
\begin{equation}
    \delta_{\rm app} = \frac{1}{\Gamma (1-\beta~{\rm cos}~\theta)}
\end{equation}    
and
\begin{equation}    
    \delta_{\rm cj} = \frac{1}{\Gamma (1+\beta~{\rm cos}~\theta)},
\end{equation}
where $\Gamma$ is the Lorentz factor for the jet, $\beta$ is the ratio of the jet speed to the speed of light and $\theta$ is the angle between the jet axis and the line of sight.  For jets where the jet axis is aligned with the angular momentum axis of the orbit, $\theta$ will be equal to the binary inclination angle, although alignment is not necessary for our conclusions to hold.

At a given height along the jet axis, the rest frame wavelength of the emission along the counterjet will be longer than that rest frame wavelength of the emission in the approaching jet by $\delta_{\rm ratio}$, where:
\begin{equation}
    \delta_{\rm ratio} = \frac{\delta_{\rm app}}{\delta_{\rm cj}} = \frac{(1+\beta~{\rm cos}~\theta)}{(1-\beta~{\rm cos}~\theta)}.
\end{equation}
Given the relation between wavelength and height along the jet, then, $h'$ will be smaller than $h$ by this same factor.

Next, we can consider the relevant lags in the problem.  There will be light travel time delays and jet propagation delays.  The light travel time delays come from the fact that the emission from the approaching jet has a shorter distance to travel to the observer than the emission from the counterjet.  For the approaching jet, the emission will have $\Delta x = h {\rm cos}~\theta$ less distance to travel than light from the central compact object, while for the counterjet, the emission will have $\Delta x' = h' {\rm cos}~\theta$ more distance to travel.  This distance will be travelled at the speed of light and the sum of the distances will be:
\begin{equation}
    \left(\frac{h}{c}\right){\rm cos}~\theta \left (1+\frac{1}{\delta_{\rm ratio}}\right).
\label{travellag}
\end{equation}
For the jet propagation timescales, the difference, rather than the sum, is relevant.  This difference will be:
\begin{equation}
\frac{h'}{\beta c}-\frac{h}{\beta c}=\left(\frac{h}{\beta c}\right)\left(\frac{1}{\delta_{\rm ratio}}-1\right),
\label{jetlag}
\end{equation}
and by writing the equation in this order, we can sum the results from equations \ref{travellag} and \ref{jetlag}, and obtain the final total lag:
\begin{equation}
    \left(\frac{h}{c}\right)\left[\left(1+\frac{1}{\delta_{\rm ratio}}\right){\rm cos}~\theta +\left(\frac{1}{\beta}\right)\left(\frac{1}{\delta_{\rm ratio}}-1\right)\right].
\end{equation}
Factoring the terms in the denominators inside the parentheses, we get:
\begin{equation}
    \left(\frac{h}{\delta_{\rm ratio}\beta c}\right)\left[(1+\delta_{\rm ratio})\beta {\rm cos}~\theta +\left({1-\delta_{\rm ratio}}\right)\right].
\end{equation}

Then, we can write out $\delta_{\rm ratio}$, and obtain:
\begin{equation}
    \left(\frac{h}{\delta_{\rm \rm ratio}\beta c}\right)\left[\left(1+\frac{(1+\beta~{\rm cos}~\theta)}{(1-\beta~{\rm cos}~\theta)}\right)\beta {\rm cos}~\theta +\left({1-\frac{(1+\beta~{\rm cos}~\theta)}{(1-\beta~{\rm cos}~\theta)}}\right)\right].
\end{equation}
We can then multiply through by $(1-\beta~{\rm cos}~\theta)$, and it will become readily apparent that the expression is zero.

\section{Highlighting the assumptions and applicability of the above result}

Several specific assumptions have been made above, and violations of those assumptions could lead to different arrival times at the observer for the approaching and counterjets.  One of the more important assumptions is that the jet is powered symmetrically.  If the jet is not powered symmetrically, then the light curves of the approaching and counterjets should be statistically similar, but independent of one another.  Statistically, the delays from the accretion emission to the jet emission would be the same for the approaching jet and counterjet.  

Non-steady jets sometimes appear to be consistent with being symmetric \citep{Mirabel, Tetarenko2017}, and in other cases appear to have some inherent asymmetry \citep{HjellmingR.M1995Eeor}.  It may be possible though, that even the case of GRO~J1655--40 studied by \citet{HjellmingR.M1995Eeor} is symmetric, but with larger swings in inclination angle than considered in that work, more similar to the large swings seen in V404 Cyg \citep{Tetarenko2017}.

Next, it has been assumed that the jet speed is constant.  From studies of the few active galactic nuclei with well-resolved jets, it is clear that they are being accelerated as they move out from the central black hole, even within the region in which they emit strongly (e.g. \citealt{Lister2021}).  This can lead to a non-zero counterjet lag behind the approaching jet, while conversely deceleration can lead to the counterjet emission arriving first.  

\section{Testing the twin jet hypothesis}

A clear model test exists for whether the jets are symmetric in the steady states.  Many theoretical calculations suggest that the jets should be asymmetric for short amounts of time (e.g. \citealt{McKinney2012}, as well as some of the simulations released via the Event Horizon Telecope collaboration\footnote{https://www.youtube.com/watch?v=1Sv7djCASDg{\&}t=1s}).  Anecdotally, it is well known that general relativistic magnetohydrodynamic simulations often produce short-term asymmetries in jet power, but this effect has not been systematically studied in simulations (P.C. Fragile, private communication).

If two independent jet components, with largely similar power spectra are being summed, this should not affect the statistical properties of the power spectrum at all if there is no systematic lag between them.  It {\it will} lead to a loss of cross-coherence\footnote{This is often referred to as just the coherence, but as coherence is sometimes used to refer to the quality factor of periodic oscillations, we prefer to use the term cross-coherence to avoid confusion.} between the radio and X-ray bands -- see \citet{Vaughan1997} for a discussion of the cross-coherence, if calculated on sufficiently short timescales and for a source for which the counterjet represents a sufficiently large faction of the flux. In the context of the \citet{Malzac2018} model, the approaching and counterjets should be identical, and this model does a good job of explaining the observations.  Still, given the lack of cross-coherence, it is possible that the approaching jet responds primarily to the ``top" half of the accretion flow, and the receding jet responds to the ``bottom" half, then the two components could have statistically identical, but full independent correlations.  This could take place in the context of a model that preserves the successful features of the \citet{Malzac2018} work.

In this case, if we take $\alpha$ as the ratio of the counterjet's flux to that of the approaching jet, we find that the expected cross-coherence of the summed jet emission, $\gamma_j^2$ should be:
\begin{equation}
    \gamma^2_j=\gamma^2\left(\frac{1+\alpha^2}{1+2\alpha+\alpha^2}\right),
    \label{coherencesummed}
\end{equation}
with the derivation of this result shown in Appendix A.
For $\alpha$=1, corresponding to a 90 degree inclination angle for the jet, the cross-coherence should be reduced by a factor of 2.  For fast, pole-on jets, the cross-coherence should be, unsurprisingly, nearly unaffected by the presence of the counterjet emission.  For neutron star jets, then, where $\alpha$ can be in the 0.1-0.5 range for typical speeds and inclination angles, it is reasonable to expect that (1) the cross-coherence will be weaker than it is for black hole jets and (2) that it will be lower for slower jets and for jets closer to perpendicular to the line of sight than it is for jets that are pole-on, or are at the faster end of the range of parameters for accretion neutron stars.  These quantities have not yet been well measured for rapid variability from neutron star X-ray binaries as sensitive radio data sets with high time resolution have not yet been obtained for accreting neutron star X-ray binaries.  The most likely candidate among the known stellar mass black holes for showing effects from its counterjet is GRO~J1655-40, which has shown outbursts in 1994, 1996, and 2005 \citep{1655outburst,WATCHDOG}, and for which the best estimate of the jet inclination angle is 85 degrees \citep{HjellmingR.M1995Eeor}.  This system, too, should have relatively similar fluxes from the approaching jet and counterjet.

Given the broad similarities of neutron star jets to those of black hole X-ray binary jets, we can draw intuition for what to expect from the black hole systems.  For black hole X-ray binaries, there are good cross-coherence measurements between infrared and X-ray emission (e.g. \citealt{Vincentelli}), for GX~339-4.  The infrared band for typical black hole X-ray binaries will be only from the approaching jet, because the counterjet in the infrared will be behind the optically thick part of the accretion disk at the heights where the infrared emission is produced \citep{Maccarone2020}.  \citet{Tetarenko2021} does not report cross-coherences between radio and X-rays explicitly, but given that the normalizations of the cross-correlation functions between X-ray and radio are typically $\sim0.5$, it is likely that the cross-coherence is of the same order over the frequency range at which the variability is strongest.  A substantial fraction of the neutron star X-ray binaries, primarily the ``Z-sources" have sufficient radio flux densities to perform rapid variability analyses with current instrumentation (noting that only at certain positions along the Z-track do they show flat spectra at centimeter wavelengths), as do a few of the atoll sources, but a comprehensive study of the rapid variability of neutron star jets is likely to require the Next Generation Very Large Array (ngVLA; \citealt{ngVLA}).  A clear prediction can be made that if the high Fourier frequency cross-coherence between radio and X-ray emission is a strong function of binary inclination angle, then the approaching and counterjets are likely to be independent of one another, while if there is no inclination angle dependence, the ``twin jet" model genuinely applies on short timescales.

\section{Challenges in looking for jet acceleration}

\begin{figure*}
\includegraphics[width=0.6\textwidth]{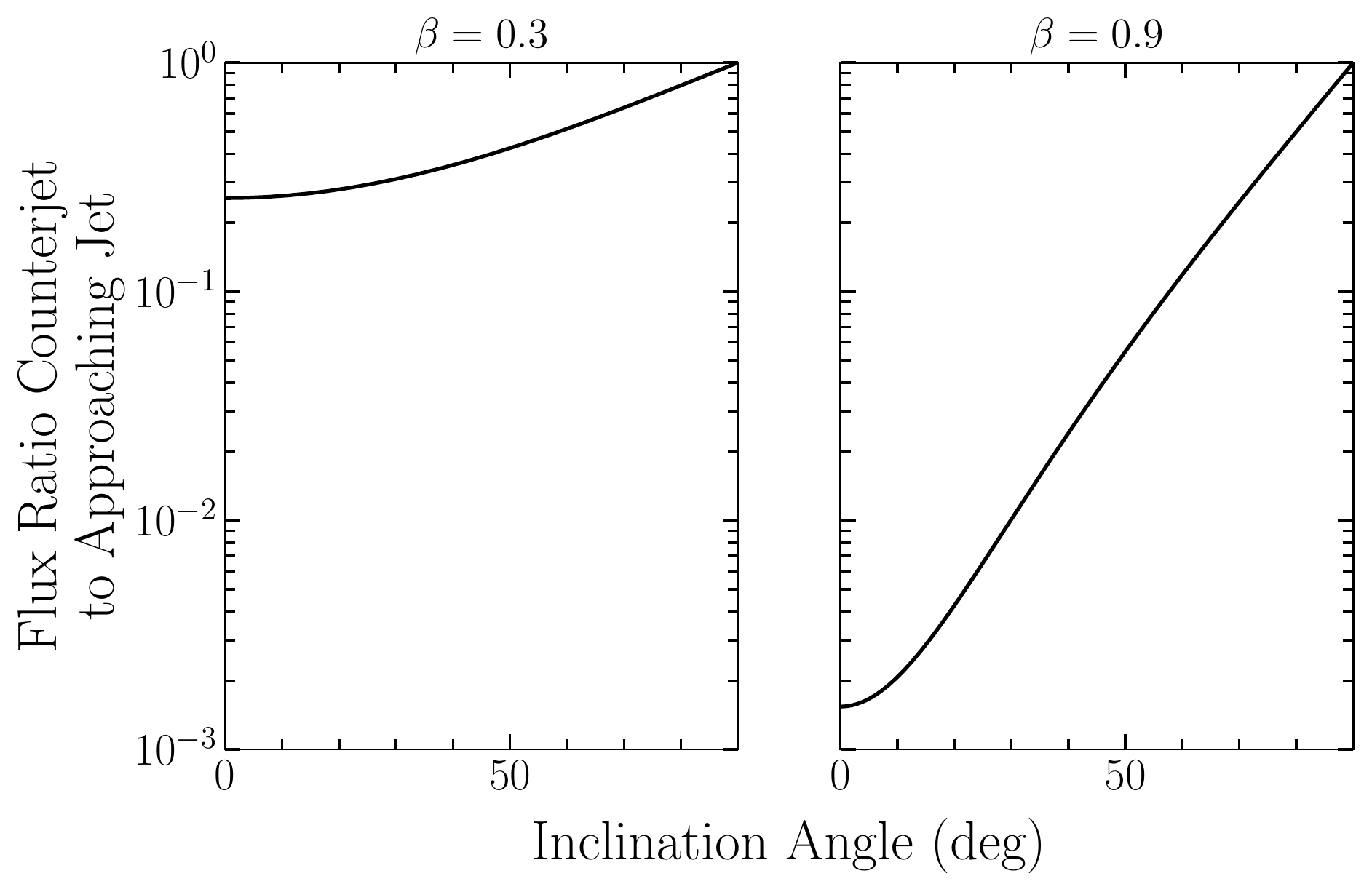}
\caption{The ratio of the flux for the counterjet to that for the approaching jet as a function of inclination angle.  Left: for $\beta=0.3$.  Right: for $\beta=0.9$. { We assume that the flux will scale as the ratio of the Doppler factors to the 2.2 power \citep{2016MNRAS.463.1153Z}}. }
\label{ratiofig}
\end{figure*}

Two major challenges exist in looking for the signatures of jet acceleration by using counterjets.  One is that Doppler boosting makes the counterjets much fainter than the approaching jets, especially for speeds $\beta \gtrsim 0.9$, as shown in figure \ref{ratiofig}.  
From this, it is clear that only for relatively slow jets, or relatively high inclination angles is there a substantial fraction of the flux from the counterjet.

The other is that the time lag difference for the approaching jet and counterjet is likely to be at least ten times shorter than the lag for the approaching jet in cases where there is modest acceleration.  Empirically, the smearing of the signature of the jet gives a break timescale in the jet power spectrum that is comparable to the time lag for the approaching jet \citep{Tetarenko2021}.  This smearing occurs mostly because the \citealt{BK79} model leads to the jet emission at a given wavelength coming from a range of heights along the jet axis.  Because of the self-similarity of the jets, the result above about the lag timescales being the same for approaching and counterjets with constant speeds still holds.

A mathematical technique called the `cepstrum' has been developed \citep{Bogert:1963:FAT} to search for echoes in light curves.  The cepstrum is the power spectrum (produced from an inverse Fourier transform) of the logarithm of the power spectrum (produced using a forward Fourier transform).  In many cases, a cepstrum of a signal added to its own echo will yield a delta-function, with the time lag associated with the echo.  This can be seen following the work of \citep{oppenheim2004frequency}.

One can take:
\begin{equation}
    x(t)=s(t)+\alpha s(t-\tau),
\end{equation}
where $s(t)$ is the initial signal, and $\tau$ is the timescale of the echo.
The, the power spectrum $|X(f)|^2$ will be given by:
\begin{equation}
    |X(f)|^2 = |S(f)|^2\left[1+\alpha^2+2\alpha{\rm cos}(2\pi f\tau)\right].
\end{equation}
The logarithm then converts the multiplication into a sum of the log of the power spectrum of the underlying signal times the term inside the square brackets.  The inverse Fourier transform power spectrum then pulls out the period of the cosine term.

An important caveat remains.  The second maximum of the cosine will occur at $f=\frac{1}{\tau}$.  If $|S(f)|^2$ is extremely small at this frequency, then the power spectrum $|X(f)|^2$ of the summed signal will be dominated there by Poisson noise.  The cepstrum is thus sensitive only for cases where the echo timescale is slow relative to the characteristic timescale of the variability.  Because, as mentioned above, the break frequency in the power spectrum is typically comparable to the reciprocal of the time lag, only with either exquisite signal-to-noise, or with lags between the approaching and counterjets that are large fractions of the lags of the approaching jet behind the disk emission, can we expect to detect the counterjets via this approach.

There appear to be realistic scenarios in which the cepstral lags could be measured, provided that the approaching jet reaches a relatively high speed ($\Gamma$ of at least about 5).  The large speed is needed to ensure that the doppler factors for the approaching jet and counterjet are sufficiently different that the precise cancellation for the non-accelerated jet starts to fail substantially.

If we take the case, for example, of a jet which accelerates with constant acceleration from $\beta=0.7$ at its base to $\beta=0.98$ at the region of interest for emission of the approaching jet, and take an inclination angle of 60 degrees the same wavelength will come from a region with $\beta=0.91$, only about 70\% the distance to the black hole.  It has been shown for blazar jets that there cannot be very high speeds on very small spatial scales, or Compton drag would slow the jets down, so the general picture of whether such accelerations take place is well worth testing \citep{BegelmanSikora}, and strong empirical evidence exists for acceleration of blazar jets in VLBI data \citep{Lister2021}, but due to the long variability timescales of AGN jets, studies of them are inherently nonstationary.  In such a scenario, the jet to counterjet flux ratio for relative edge-on jets would be mitigated, both because (1) both jets are deboosted and (2) the counterjet's speed is slower at the relevant wavelength, so the deboosting ratio is mild.  For the scenario laid out above, the jet to counterjet flux ratio should only be a factor of about 3.  Cepstral searches of existing data for such objects are thus well-motivated, and long, high time resolution data sets in radio through infrared are well-motivated for future outbursts because of this.  One potential complicating factor of which future observers should be mindful is that a delay may also be present in some systems in relatively short wavelength bands like the optical and near infrared due to thermal reprocessing in the accretion disc (and, in many systems, due to the fact that the accretion disc itself will block the inner parts of the counterjets that produce emission in these bands).

The effects on the cross-coherence of the jets will be more complicated in the situation where there is jet acceleration that is relevant.  In such a case, the counterjet will be the sum of two distinct emission regions, with a fixed relation between the regions.  Following \citet{Vaughan1997}, equation 10, this will give an intermediate cross-coherence between the two extremes.

\section{Summary and conclusions}

We have found several key results from this work:
\begin{enumerate}
    \item For the standard assumptions of a constant speed jet following the \citet{BK79} model, the emission at a given wavelength from the approaching jet and the receding jet will arrive to a distant observer simultaneously.  
    \item In that scenario, one can test whether the jet is genuinely symmetric by looking at the cross-coherence between X-ray emission and emission from some band produced by the jet.
    \item In the case of jet which is accelerating within the emission region at a given wavelength, there can be time lags between the approaching and receding jets that should be measurable using the cepstrum.
    \item In both cases (ii) and (iii) above, the measurements should be most valuable in the highest frequency bands for which the counterjet is not blocked by the outer accretion disk.  This will typically be the far-infrared or submillimetre band, but as the ngVLA project \citep{ngVLA} starts to collect data, its superior sensitivity may make it the instrument of choice.  { One core change that may be necessary relative to current observational set-ups is that longer data sets should be obtained to make precise measurements of the coherence than are needed to estimate the lags.}
\end{enumerate}

\section{Acknowledgments}
We thank the anonymous referee for a helpful report.  We thank Sara Motta, James Miller-Jones, Greg Sivakoff, Piergiorgio Casella for useful discussions.  We also thank the participants of a workshop in honour of Omer Blaes' 62nd and Chris Fragile's 52nd birthdays (and particularly the honourees) for useful discussions. Support for this work was provided by NASA through the NASA Hubble Fellowship grant \#HST--HF2--51494.001 awarded by the Space Telescope Science Institute, which is operated by the Association of Universities for Research in Astronomy, Inc., for NASA, under contract NAS5--26555.

\section{Data availability}
No new data were collected for the work presented here.



\bibliographystyle{mnras}
\bibliography{bibliography}

\appendix 
\section{Coherence of two summed signals}

To evaluate the expected cross-coherence from adding together the emission from two jet components, we start with a disk component, $d(t)$ and two jet components, $j_1(t)$ and $j_2(t)$.  The disk component will be measured from X-rays, while the jet components will be measured from two different locations, but using the same band, so that the total jet emission is $j(t)$, and with angular resolution of current instruments, these cannot be separated.

It is already established that the cross-coherence of the jet emission is well below unity, so we start by developing an expression for the coherent and incoherent parts of the jet emission:

\begin{equation}
    \gamma^2=\frac{|<D^*J_1>|^2}{<|D|^2><|J_1|>^2},
\end{equation}
where $D$ denotes the Fourier transform of $d(t)$, $J_1$ denotes the Fourier transform of $j_1(t)$, the asterisk denotes a complex conjugate, and the $\gamma^2$ is the cross-coherence.  The Fourier components and the cross-coherence are all functions of frequency, but this will be implicit in the equations in this section in the interests of legibility.

Next, we consider that each direction's jet component is the sum of a correlated component and an uncorrelated component:
\begin{equation}
<D^*J_1> = <D^*(J_{1,{\rm corr}}+J_{1,{\rm uncorr}})>,    
\end{equation}
where $J_{\rm corr}$ and $J_{\rm uncorr}$ are the correlated and uncorrelated components of the Fourier transform, respectively.

Then, when we expand the numerator, the term with the uncorrelated component will tend to zero under the time averaging, so we can find that:
\begin{equation}
    |<D^*J_{\rm corr}>|^2=\gamma^2<|D|^2><|J_1|^2>.
\end{equation}

Next, we consider the sums of the two jet components.  Here, we wish to compute $\gamma_j^2$, which should be found from:
\begin{equation}
    \gamma_j^2=\gamma^2 \frac{|<D^*(J_1+J_2)>|^2}{<|D|^2><|(J_1+J_2)|^2>}.
\end{equation}

In the limiting case where $J_1$ and $J_2$ are the same time series, then the numerator and denominator of the fractional term on the right cancel, and $\gamma_j^2$=$\gamma^2$.  

In the limiting case where the two components are completely independent, albeit both with the same cross-coherence relative to the disk emission, we can follow the derivation in equation 10 of \citep{Vaughan1997}, and we find the result in equation \ref{coherencesummed} of this paper.

\bsp	

\label{lastpage}
\end{document}